\newcommand{\vece}{\mbox{\boldmath$e$}}

\newcommand{\vecj}{\mbox{\boldmath$j$}}

\newcommand{\vecs}{\mbox{\boldmath$s$}}
\newcommand{\vecr}{\mbox{\boldmath$r$}}

\newcommand{\vecv}{\mbox{\boldmath$v$}}

\newcommand{\vecA}{\mbox{\boldmath$A$}}
\newcommand{\vecB}{\mbox{\boldmath$B$}}

\newcommand{\vecE}{\mbox{\boldmath$E$}}

\newcommand{\dfd}{{\rm d}}

\newcommand{\lagr}{{\cal L}}
\newcommand{\hami}{{\cal H}}
\newcommand{\arcsinh}{\mbox{ arcsinh}}

\newcommand{\half}{\frac{1}{2}}
\newcommand{\emf}{e.m.f.}

\newcommand{\lwp}{Li\'enard-Wiechert potentials}
\newcommand{\ie}{{\em i.e.}}
\newcommand{\eg}{{\em e.g.}}
\newcommand{\etal}{{\em et al.}}

\documentclass[10pt,a4paper]{article}
\usepackage{latexsym}
\usepackage[]{amssymb}
\usepackage[]{graphicx}
\usepackage[]{epsfig}

\title{From least action in electrodynamics to
magnetomechanical energy -- a review}
\author{Hanno Ess\'en\\
Department of Mechanics \\Royal Institute of Technology  \\ SE-100
44 Stockholm, Sweden}
\date{January, 2009}

\begin{document}

\maketitle

\begin{abstract}
The equations of motion for electromechanical systems are traced
back to the fundamental Lagrangian of particles and
electromagnetic fields, via the Darwin Lagrangian. When
dissipative forces can be neglected the systems are conservative
and one can study them in a Hamiltonian formalism. The central
concepts of generalized capacitance and inductance coefficients
are introduced and explained. The problem of gauge independence of
self-inductance is considered. Our main interest is in
magnetomechanics, \ie\ the study of systems where there is
exchange between mechanical and magnetic energy. This throws light
on the concept of magnetic energy, which according to the
literature has confusing and peculiar properties. We apply the
theory to a few simple examples: the extension of a circular
current loop, the force between parallel wires, interacting
circular current loops, and the rail gun. These show that the
Hamiltonian, phase space, form of magnetic energy has the usual
property that an equilibrium configuration corresponds to an
energy minimum.
\end{abstract}

\newpage
\section{Introduction}
Electromagnetism is usually taught at the undergraduate level
without mention of Lagrangians, Hamiltonians, or the principle of
least action. In modern theoretical physics of gauge field theory,
however, the concept of an invariant Lagrangian density has become
the standard starting point. The Lagrangian formalism of
analytical mechanics was introduced into electromagnetism already
by Maxwell in his Treatise \cite{BKmaxwell} who, using this
approach, derives equations for electric circuits and for
electromechanical systems. Since then its importance has kept
growing. One can therefore argue that this set of tools should be
better known and become accessible at an earlier stage in the
physics curricula. This review attempts to be an aid in such
efforts.

Our starting point is the basic Lagrangian density of classical
electrodynamics as set down early in the last century by Larmor
and Schwarzschild. From there we proceed to neglect radiation
which leads us to the Darwin Lagrangian \cite{darwin}. Then the
path to classical linear circuit theory is traced. It is pointed
out that the similarity between the Lagrangian formulation of
mechanics and of circuit theory has deep physical reasons and is
not just a formal similarity. In a long Appendix the generalized
capacitance coefficients and the coefficients of self and mutual
induction of circuit theory are derived, investigated and
explained.

Lagrangians of electromechanical systems are also seen to arise
from the Darwin Lagrangian by introducing suitable constraints, or
assumptions, on the possible movements of both the charged
particles and the neutral matter in the system. We concentrate on
magnetomechanical problems, \ie\ problems where there is a
magnetic interaction energy involving macroscopic matter. As
examples of such problems we consider the extension of a circular
loop of current, the attraction of parallel currents, the
interaction between two circular loops of current, and the rail
gun. Finally we discuss the properties of the concept of magnetic
energy and clarify some tricky points.

\section{Lagrangian electrodynamics} In modern physics one has
found that the most reliable and fundamental  starting point in
theoretical investigations is the principle of least action. The
action is a scalar quantity constructed from a Lagrange density
which is a function of the relevant particle and field variables
and their (normally first) derivatives. The action for classical
electrodynamics is the time integral of the Lagrangian $\lagr$
which has three parts,
\begin{equation}\label{eq.tot.L.three.parts}
\lagr =\lagr_{\rm m}+\lagr_{\rm i}+\lagr_{\rm f}.
\end{equation}
The first part is the Lagrangian for free non-interacting
particles,
\begin{equation}
\label{eq.rel.kin.lagr} \lagr_{\rm m} = \sum_{a=1}^N
 \lagr_{{\rm m}\,a}= \sum_{a=1}^N -m_a
c^2\sqrt{1-\vecv_a^2/c^2}.
\end{equation}
In the non-relativistic approximation, which we mostly assume
valid, it is simply the kinetic energy. The second is the the
interaction Lagrangian,
\begin{equation}\label{eq.L.interact1}
\lagr_{\rm i} =\int \left( \frac{1}{c} \vecj\cdot\vecA -
\varrho\phi \right) \dfd V .
\end{equation}
It was published in 1903 by Karl Schwarzschild (1873 - 1916) and
describes the interaction of the charge and current density of the
particles with the electromagnetic potentials. The third and final
part is the field Lagrangian,
\begin{equation}\label{eq.field.lagr}
\lagr_{\rm f} =\frac{1}{8\pi}\int (\vecE^2 -\vecB^2)\,\dfd V ,
\end{equation}
originally suggested by Joseph Larmor (1857 - 1942) in 1900. The
connection with (\ref{eq.L.interact1}) is via the identifications,
\begin{equation}\label{eq.field.from.pots}
\vecE=-\nabla \phi -\frac{1}{c}\frac{\partial \vecA}{\partial
t},\; \mbox{and}\;\; \vecB= \nabla\times\vecA.
\end{equation}
Maxwell's homogeneous equations are identities obtained by taking
the curl of the first of the equations (\ref{eq.field.from.pots}),
and the divergence of the second. Maxwell's remaining,
inhomogeneous equations, and the equations of motion for the
particles under the Lorentz force, are all obtained from the
variation of the action, $S = \int \lagr \dfd t$, with $\lagr$
from (\ref{eq.tot.L.three.parts}). It is this joining of both the
equations determining the fields from the sources, and the
equations of motion of the sources due to the fields, into a
single formalism, that is the strength and beauty of this
approach.

The variational approach to electromagnetism outlined above can be
found in many of the more advanced textbooks on electrodynamics
\cite{BKjackson3,BKlandau2,BKkonopinski,BKschwinger&al,BKpanofsky,BKlanczos}.
More specialized works are Yourgrau and Mandlestam
\cite{BKyourgrau}, Doughty \cite{BKdoughty}, and Kosyakov
\cite{BKkosyakov}.

\subsection{The Darwin Lagrangian}
In many types of problems one can neglect the radiation of
electromagnetic waves from the system under study, since this
phenomenon is proportional to $c^{-3}$. In those circumstances the
field Lagrangian $\lagr_{\rm f}$ can be rewritten and one finds
that $\lagr_{\rm f} = -\half \lagr_{\rm i}$. Inserting this in
(\ref{eq.tot.L.three.parts}) we get,
\begin{equation}\label{eq.action.no.rad}
\lagr=\lagr_{\rm m} + \half \lagr_{\rm i} ,
\end{equation}
for the relevant Lagrangian in the non-radiative case. When the
motion of a charged particle is known one can find the potentials,
$\phi, \vecA$, that it produces, the so called retarded, or \lwp.
Expanding these to order $(v/c)^2$, one finds that acceleration
vanishes from the Lagrangian (since it only contributes a total
time derivative to this order). The result is a Lagrangian that
contains only particle positions and velocities. There are then no
independent electromagnetic field degrees-of-freedom. Everything
is determined by the positions and velocities of the charged
particles, and the resulting Lagrangian is the Darwin Lagrangian
\cite{darwin}, as derived by Charles Galton Darwin (1887 - 1962),
a grandson of the great naturalist, in 1920.

The Darwin Lagrangian can be written,
\begin{equation}\label{eq.darw.lagr.density.form}
\lagr_D = \lagr_{\rm m} + \half \int \left( \frac{1}{c}
\vecj\cdot\vecA - \varrho\phi \right) \dfd V ,
\end{equation}
\ie\ Eq.\ (\ref{eq.action.no.rad}), where,
\begin{equation}\label{eq.phi.in.terms.of.rho}
\phi(\vecr) =\int \frac{\varrho(\vecr')\,\dfd V'}{|\vecr-\vecr'|},
\end{equation}
which is exact in the Coulomb gauge, and where,
\begin{equation}\label{eq.darwin.vec.pot.curr.dens.int}
\vecA(\vecr) = \frac{1}{2c} \int  \frac{\vecj(\vecr') +
[\vecj(\vecr')\cdot \vece_{r' r}]\vece_{r' r} }{ |\vecr -
\vecr'|}\,\dfd V'.
\end{equation}
Here $\vece_{r' r} =(\vecr - \vecr')/|\vecr - \vecr'|$. This
specific form of the Darwin vector potential can be traced back to
a fairly large retardation effect in the Lorenz gauge Coulomb
potential. Its effect is included in the Darwin approximation
which, however, uses a Coulomb gauge, $\nabla\cdot\vecA = 0$.

A more familiar form of the Darwin Lagrangian, for $N$ point
particles, is obtained by introducing,
\begin{equation}\label{eq.curr.dens.points}
\varrho(\vecr)= \sum_{a=1}^N  e_a\, \delta\!\left(\vecr
-\vecr_a\!(t)\right),\; \mbox{and}\;\;\vecj(\vecr) = \sum_{a=1}^N
e_a \vecv_a\!(t)\, \delta\!\left(\vecr -\vecr_a(t)\right) ,
\end{equation}
in the expressions (\ref{eq.darw.lagr.density.form}) -
(\ref{eq.darwin.vec.pot.curr.dens.int}) given above. After
skipping self interactions one obtains,
\begin{eqnarray} \label{eq.LtotNoRad2} \lagr_D = \lagr_{\rm m}
+ \half \sum_{a=1}^N  \left( \frac{1}{c} e_a \vecv_a \cdot \vecA_a
(\vecr_a) -
 e_a\phi_a(\vecr_a)  \right) ,\\
\nonumber {\rm where,} \\
\label{eq.coul.pot} \phi_a (\vecr) = \sum_{b(\neq a)}^N
\frac{e_b}{|\vecr -\vecr_b|}, \\
\nonumber {\rm and,} \\
\label{eq.darwin.A.ito.velocity} \vecA_a (\vecr) = \sum_{b(\neq
a)}^N \frac{e_b [\vecv_b + (\vecv_b\cdot\vece_{r_b r}) \vece_{r_b
r} ] }{2c|\vecr-\vecr_b|} .
\end{eqnarray}
Here $\vecr_a$ and $\vecv_a$ are particle position and velocity
vectors respectively, $m_a$ and $e_a$ their rest masses and
charges respectively, while $\vece_{r_b r} =
(\vecr-\vecr_b)/|\vecr-\vecr_b|$. There are no independent field
degrees-of-freedom and hence no gauge invariance in the Darwin
formalism, which entails action-at-a-distance. Retardation is
included to order $(v/c)^2$, a fact which is often missed in the
literature.

The Darwin approach to electromagnetism is only briefly mentioned
in some advanced textbooks
\cite{BKjackson3,BKlandau2,BKschwinger&al}. A book by Podolsky and
Kunz \cite{BKpodolsky} is a bit more thorough. Several good
fundamental and pedagogical studies can, however, be found in the
literature
\cite{anderson67,barker,breitenberger,essen96,essen99,kennedy,krause2,larsson}.

\subsection{The kinetic energy of currents}
In the free particle Lagrangian $\lagr_{\rm m}$ of Eq.\
(\ref{eq.rel.kin.lagr}) the approximation,
\begin{equation}\label{eq.approx.lagr.kin}
\lagr_{{\rm m}\,a} \approx \displaystyle - m_a c^2 +\frac{m}{2}
\vecv_a^2 + \frac{m_a}{8c^2}\vecv_a^4,
\end{equation}
is usually done, because of the validity of the Darwin approach to
order $(v/c)^2$. Here we will be concerned with systems in which
there are macroscopic charge and current densities confined to
electrically conducting matter. In 1936 Darwin \cite{darwin2}
found that the magnetic energy contribution to the inertia of the
conduction electrons is roughly $10^8$ greater that the
contribution from their rest mass. This means that, so called,
inductive inertia dominates. For the dynamics of macroscopic
currents and charges in fixed conductors (electric circuit theory)
one can consequently also neglect the free particle Lagrangian
$\lagr_{\rm m}$. In plasma physics the neglect of particle inertia
is called the force free approximation \cite{lundquist}.

Skipping $\lagr_{\rm m}$,
\begin{equation}\label{eq.lagr.negle.rest.mass}
\lagr_{\rm LC} = \half \int \left( \frac{1}{c} \vecj\cdot\vecA -
\varrho\phi \right) \dfd V ,
\end{equation}
is all that then remains of (\ref{eq.darw.lagr.density.form}).
This Lagrangian, together with $\vecA$ and $\phi$ given by
(\ref{eq.phi.in.terms.of.rho}) and
(\ref{eq.darwin.vec.pot.curr.dens.int}) respectively, describes
electromagnetic systems with inductive and capacitive phenomena.
For electromechanical systems, on the other hand, the
non-relativistic form of kinetic energy must be retained for the
mechanical degrees-of-freedom. Potential energy contributions due
to elasticity or gravitation may also have to be included.

\section{Linear electric circuits} The equations governing
linear electric circuits are presented in almost every textbook on
electromagnetism, and their similarity with those for oscillating
mechanical systems is often pointed out. A smaller number of more
advanced texts even go as far as presenting a Lagrangian formalism
underlying the circuit equations
\cite{BKguillemin,BKlandau8,BKvagner}.

Here we will derive and discuss some standard results in for
linear electric circuits starting directly from
(\ref{eq.lagr.negle.rest.mass}). These are alternatively called
current circuits, or networks, in the literature. Assume that all
current flows in conducting thin (filamentary) wires and that
there are $n$ such wires with currents $i_k =\dot e_k,\;
(k=1,\ldots,n)$. It is then easy to show that the magnetic part of
(\ref{eq.lagr.negle.rest.mass}) can be written,
\begin{equation}\label{eq.lagr.syst.wire.currents}
\lagr_{\rm L} =\half \int  \frac{1}{c} \vecj\cdot\vecA\, \dfd V =
\half \sum_{k=1}^n \sum_{l=1}^n  L_{kl}  \dot e_k \dot e_l .
\end{equation}
In a similar way for a fixed arrangement of $m$ (extended)
conductors, with charges $e_i\; (i=1,\ldots,m)$ on them, the
electric part of $\lagr_{LC}$ can be written,
\begin{equation}\label{eq.lagr.syst.charged.conduct}
\lagr_{\rm C} =-\half \int  \varrho\phi\,
 \dfd V =- \half \sum_{i=1}^m \sum_{j=1}^m  \Gamma_{ij} e_i e_j .
\end{equation}
The inductance coefficients $L_{kl}$ and the generalized
capacitance coefficients $\Gamma_{ij}$ only depend on the geometry
of the arrangement. These are derived and explained in the
Appendices. We have thus found that the Lagrangian
(\ref{eq.lagr.negle.rest.mass}) under the above assumptions can be
written,
\begin{equation}\label{eq.lagr.syst.conduct.plus.cond.wires}
\lagr_{\rm LC}(e, \dot e) = \lagr_{\rm L}(\dot e) + \lagr_{\rm
C}(e).
\end{equation}
This is a valid total Lagrangian for a non-radiating arrangement
of current carrying thin wires and extended charged conductors.
This separation of magnetic and electrostatic effects comes from
the central idea that there will be no net charge density on thin
wires, and that currents in extended conductors have negligible
magnetic effects.

\subsection{The conductor pair condenser}
\begin{figure}[h]
\centering
\includegraphics[width=230pt]{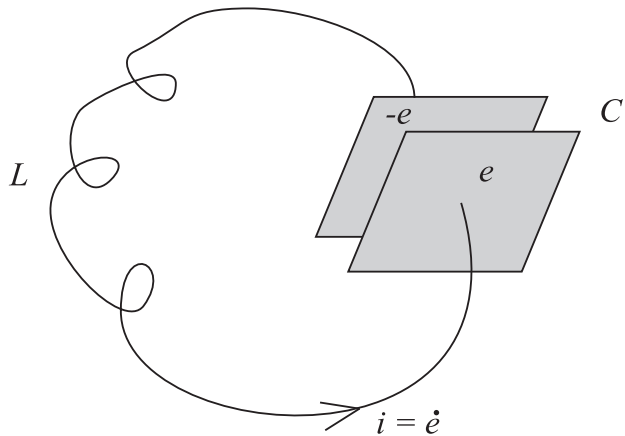}
\caption{\small A single $LC$-circuit with an inductance $L$
and  a capacitance $C$. The Lagrangian
(\ref{eq.lagr.lin.electric.circuit}) gives the dynamics for
$n$ interacting circuits of this type. \label{fig.LCcircuit}}
\end{figure}

In practice all the charges $e_i$ on all the $m$ different
conductors are not independent. Often a circuit is arranged so
that the conductors come in pairs that are very close, so called
condensers. If each such pair is connected by a wire while being
electrically isolated otherwise, the total charge on that
subsystem must be a constant which we take to be zero. The number
of wires is then half the number of conductors, $n=\frac{m}{2}$,
and the charges $e_i$ come in pairs that are equal and opposite
$e_k = -e_{k+n}, (k=1,\ldots,n=\frac{m}{2})$, while the current in
the wire connecting them is $i_k = \dot e_k$, see Fig.\
\ref{fig.LCcircuit}. There is then only $n=\frac{m}{2}$
degrees-of-freedom of the problem. We now assume, without loss of
generality, that the coefficients $\Gamma_{ij}$ are symmetric in
the indices $ij$, and define the new symmetric matrix,
\begin{equation}
\label{eq.new.def.of.capac}
 C^{-1}_{kl} = C^{-1}_{lk} \equiv \Gamma_{k\, l} +
\Gamma_{k+n\, l+n} -\Gamma_{k\, l+n} - \Gamma_{l\, k+n},
\end{equation}
where $k,l=1,\ldots,n$. Using this our Lagrangian
(\ref{eq.lagr.syst.conduct.plus.cond.wires}) can be written,
\begin{equation}\label{eq.lagr.lin.electric.circuit}
\lagr_{\rm LC}(e,\dot e) = \half \sum_{k=1}^n \sum_{l=1}^n \left(
L_{kl} \dot e_k \dot e_l - C^{-1}_{kl} e_k e_l \right).
\end{equation}
Here the $n$ charges $e_k$ and currents $\dot e_k$ are independent
and a diagonal element of the $L$ matrix, $L_{kk}$, is a
self-inductance, while the off diagonal elements correspond to
mutual inductances. A diagonal element, $C^{-1}_{kk} = \Gamma_{k\,k} +
\Gamma_{k+n\, k+n} - 2 \Gamma_{k\, k+n}$, of the $C^{-1}$-matrix represents
the inverse capacitance, $C_{kk}$, of the corresponding conductor pair
condenser.

\subsection{Equivalence of electric and mechanical oscillators}
The Lagrangian (\ref{eq.lagr.lin.electric.circuit}) is completely
equivalent to that of a mechanical system of coupled oscillators,
the $L$-matrix corresponding to the mass matrix and the
$C^{-1}$-matrix corresponding to the stiffness matrix (of spring
constants). This is often regarded as a purely formal
correspondence, a mere mathematical mapping of one problem on
another physically completely different one. This is wrong,
however. If we denote the linear density of conducting charge in
wire $k$ by $\lambda_k$, and the arc length along this wire by
$s_k$, we find that the current in the wire is, $i_k = \dot e_k =
\lambda_k \dot s_k$. Here, of course, $\dot s_k$ is the speed of
the conducting linear charge density. Clearly the charges on the
condensers have to be, $e_k = \lambda_k s_k$ (with a suitable
choice of origin and orientation for the arc length). If this is
inserted in $\lagr_{\rm LC}$ we find that,
\begin{eqnarray}
\nonumber \lagr_{\rm LC}(e, \dot e) = \half \sum_{k=1}^n
\sum_{l=1}^n \lambda_k \lambda_l \left(  L_{kl} \dot s_k \dot s_l
- C^{-1}_{kl} s_k s_l
\right) \equiv \\
\label{eq.equivalence.circuit.osc}
\\
\nonumber \half \sum_{k=1}^n \sum_{l=1}^n  \left( M_{kl} \dot s_k
\dot s_l - K_{kl} s_k s_l \right) = T(\dot s) - V(s) = \lagr(s,
\dot s).
\end{eqnarray}
The generalized coordinates $s_k$ now have dimension length so we
have an ordinary mechanical coupled oscillator Lagrangian on the
right hand side. The difference is that the mass matrix,
$M_{kl}=\lambda_k \lambda_l L_{kl}$, does not come from rest mass
but entirely from the inertia contained in the energy of the
magnetic field.

By means of the technique of simultaneous diagonalization of two
quadratic forms one can find a linear transformation to, so
called, normal mode coordinates and thus decouple the equations of
motion, which are,
\begin{equation}\label{eq.eqs.of.mot.LC.syst}
  \sum_{l=1}^n \left(  L_{kl}  \ddot e_l +
C^{-1}_{kl}  e_l \right) =0 ,\;\;\; k=1,\ldots,n,
\end{equation}
assuming that no further generalized forces enter the problem. In
terms of the normal modes $q_k$ the equations of motion become,
\begin{equation}\label{eq.eqs.of.mot.LC.syst.norm.modes}
\ddot q_k + \frac{1}{L_k C_k}  q_k  =0 ,\;\;\; k=1,\ldots,n,
\end{equation}
so these oscillate independently with angular frequencies
$\omega_k = 1/\sqrt{L_k C_k}$. For the corresponding mechanical
problem one finds $\omega_k = \sqrt{K_k/M_k}$. The expression
$\omega=1/\sqrt{LC}$ for the angular frequency of a single
$LC$-circuit, as shown in Fig.\ \ref{fig.LCcircuit}, is sometimes
referred to as Thomson's formula.

\subsection{Introduction of resistance and external voltage}
Our Lagrangian (\ref{eq.lagr.lin.electric.circuit}) corresponds to
a coupled system of undamped electromagnetic oscillators. In most
cases of practical interest the connecting wires will not be
perfectly conducting. There will be resistance in the system. The
energy will then dissipate and the equations of motion require
that there are generalized forces that describe this. Assume that
the ohmic resistance in wire $k$ is $R_{kk}$. This can be achieved
with a Rayleigh dissipation function. In a more general case there
may also be off diagonal elements $R_{kl}$, and,
\begin{equation}\label{eq.rayleigh.diss.func1}
{\cal R}(\dot e)= \half \sum_{k=1}^n \sum_{k=l}^n R_{kl} \dot e_k
\dot e_l ,
\end{equation}
is the most general form of this function for linear circuits.

If there is resistance currents $\dot e_k$ eventually dissipate to
zero and arbitrary initial conditions only lead to transient
dynamics. In most applications of circuit theory one is therefore
mainly interested in systems with added external \emf. This can be
done by the additional term,
\begin{equation}\label{eq.external.emf.lagr}
\lagr_{\rm emf}(t) = \sum_{k=1}^n e_k\, {\cal V}_k(t) ,
\end{equation}
added to the Lagrangian $\lagr_{\rm LC}$. Here ${\cal V}_k(t)$ is
an applied external voltage. A constant \emf\ will not drive a
stationary current through a condenser so to get a direct current
some of the $C^{-1}$-matrix eigenvalues must be zero. With
harmonically oscillating \emf, ${\cal V}_k(t) ={\cal V}_k \sin
(\omega t)$, capacitors are no problem and one is dealing with
alternating current circuits.

The general Lagrangian equations of motion for a system of
circuits are then,
\begin{equation}\label{eq.gen.lagrang.circuit.resist.forced}
\frac{\dfd}{\dfd t} \frac{\partial \lagr_{\rm cc}}{\partial \dot
e_k} - \frac{\partial \lagr_{\rm cc}}{\partial e_k} =
-\frac{\partial {\cal R}}{\partial \dot e_k},
\end{equation}
where,
\begin{equation}\label{eq.circuit.lagr}
\lagr_{\rm cc}(e, \dot e, t) = \lagr_{\rm LC}(e,\dot e) +
\lagr_{\rm emf}(e,t) ,
\end{equation}
is the circuit Lagrangian \cite{BKlandau8}. The system
(\ref{eq.eqs.of.mot.LC.syst}) of equations of motion are then
modified so that,
\begin{equation}\label{eq.eqs.of.mot.RLC.syst}
  \sum_{l=1}^n \left(  L_{kl}  \ddot e_l + R_{kl} \dot e_l
+ C^{-1}_{kl}  e_l \right) ={\cal V}_k(t) ,\;\;\; k=1,\ldots,n,
\end{equation}
is their new form. This is thus the type of system investigated in
linear circuit theory (see \eg\ Guillemin \cite{BKguillemin} or
Josephs \cite{BKjosephs}). In mechanical systems the ohmic
resistance terms correspond to dampers (dashpots) and the external
\emf\ to applied external force.

\subsection{Energy and Hamiltonian for conservative systems}
For Lagrangians $\lagr(q, \dot q)$ with no explicit time
dependence, such as those of Eqs.\ (\ref{eq.LtotNoRad2}) and
(\ref{eq.lagr.lin.electric.circuit}), the quantity,
\begin{equation}
\label{eq.time.deriv.prod.minus.L.integ} {\cal E}(q, \dot q)  =
\sum_{k = 1}^n \frac{\partial \lagr }{\partial \dot q_k }  \dot
q_k -\lagr ,
\end{equation}
is known to be a constant of the motion, the energy. For example
the Darwin Lagrangian (\ref{eq.LtotNoRad2}) corresponds to the
conserved energy,
\begin{equation} \label{eq.LtotNoRad2.energy} {\cal E}_D = {\cal E}_{\rm m}
+ \half \sum_{a=1}^N  \left( \frac{1}{c} e_a \vecv_a \cdot \vecA_a
(\vecr_a) + e_a\phi_a(\vecr_a)  \right).
\end{equation}
This expression for the energy goes up if currents are parallel
since the vector potential is proportional to terms like $e_b
\vecv_b$. This may seem odd since we find in Sec.\
\ref{attrac.parall.curr} that parallel currents attract. We will
return to this in Sec.\ \ref{conclus} below.

Returning to circuits we find that, when there is no time
dependent forcing and no ohmic resistance, the Lagrangian is such
that there is a conserved energy. Using
(\ref{eq.lagr.lin.electric.circuit}) and
(\ref{eq.time.deriv.prod.minus.L.integ}) gives the expression,
\begin{equation}\label{eq.energ.lin.electric.circuit}
{\cal E}_{\rm LC}(e, \dot e) = \half \sum_{k=1}^n \sum_{l=1}^n
\left(  L_{kl} \dot e_k \dot e_l + C^{-1}_{kl} e_k e_l \right),
\end{equation}
for this energy. Recall that if $\lagr=T-V$, then ${\cal E} =T+V$.
The effect of a constant forcing, due to permanent constant charge
on condensers, is only to shift the equilibrium from $e_k=\dot e_k
=0$. Ignoring this (\ref{eq.lagr.lin.electric.circuit}) is the
most general circuit Lagrangian that conserves the energy
(\ref{eq.energ.lin.electric.circuit}).

The generalized momenta obtained from the Lagrangian
(\ref{eq.lagr.lin.electric.circuit}) are by definition,
\begin{equation}\label{eq.gen.momenta.LC.circuit}
p_k \equiv  \frac{\partial \lagr_{\rm LC}}{\partial \dot e_k} =
\sum_{l=1}^n L_{kl} \dot e_l .
\end{equation}
The Hamiltonian is obtained by eliminating the generalized
velocities in the Lagrangian energy
(\ref{eq.energ.lin.electric.circuit}) in favor of the generalized
momenta. Since, $\dot e_k = \sum_{l=1}^n L^{-1}_{kl} p_l$, the
Hamiltonian will depend on the inverse of the $L$-matrix. For a
system of coupled $LC$-circuits we find the Hamiltonian,
\begin{equation}\label{eq.hamiltonian.LC.circuit}
\hami_{\rm LC}(e, p) = \half \sum_{k=1}^n \sum_{l=1}^n \left(
L^{-1}_{kl} p_k p_l + C^{-1}_{kl} e_k e_l \right),
\end{equation}
representing its conserved energy as a function of phase space
variables. The Lagrangian and Hamiltonian above, as well as the
interpretation of the canonical momenta as  magnetic fluxes
discussed below, can be found in an article by Meixner
\cite{meixner}, discussing thermodynamic issues.

\subsection{Generalized momenta and magnetic flux}
To find the meaning of the generalized, or canonical, momenta in
this case we return to the definition of the magnetic Lagrangian,
\begin{equation}\label{eq.lagr.syst.wire.currents.2}
\lagr_{\rm L} = \half \int  \frac{1}{c} \vecj\cdot\vecA\, \dfd V .
\end{equation}
The volume integration is only over the filamentary wires that
carry the currents $i_k=\dot e_k$, so, using $\vecj\,\dfd V = i_k
\dfd \vecr$, we get,
\begin{equation}\label{eq.lagr.syst.wire.currents.3}
\lagr_{\rm L} = \frac{1}{2c} \sum_{k=1}^n i_k \oint_k \vecA \cdot
\dfd \vecr ,
\end{equation}
where the line integral is around the loop of wire $k$. Now,
however,
\begin{equation}\label{eq.stokes.thoerem.for.AB}
\oint_k \vecA \cdot \dfd \vecr = \int_k (\nabla\times\vecA)\cdot
\dfd \vecs =\int_k \vecB \cdot \dfd \vecs,
\end{equation}
according to Stokes' theorem. By definition this is the magnetic
flux, $\Phi_k$, through the loop $k$. We thus find that,
\begin{equation}\label{eq.lagr.syst.wire.currents.4}
\lagr_{\rm L} = \frac{1}{2} \sum_{k=1}^n i_k \frac{\Phi_k}{c} .
\end{equation}
Comparing with (\ref{eq.lagr.syst.wire.currents}) this gives us
that,
\begin{equation}\label{eq.mag.flux.in.terms.of.inductances.eq.p}
\frac{\Phi_k}{c} = \sum_{l=1}^n L_{kl} i_l  = p_k,
\end{equation}
where the individual terms on in the sum represent contributions
to the flux through $k$ from the loops $l$ of the system.

Finally then, we have found that the generalized (canonical)
momenta $p_k$, of Eq.\ (\ref{eq.gen.momenta.LC.circuit}),
conjugate to the charges $e_k$ on the condensers are the magnetic
fluxes through the currents loops (divided by $c$): $p_k = \Phi_k
/c$. This means that if one $e_l$ does not appear in the
Lagrangian (because there is no condenser in the corresponding
loop) then that generalized momentum $p_l$ (or flux) is a constant
of the motion.

\section{Electromechanical systems}
Few texts derive of equations of motion for electromechanical
systems from the fundamental Lagrangian for particles and fields,
only Ne\u{i}mark and Fufaev \cite{BKneimark} come close. As should
be clear from the above developments electromechanical systems, as
opposed to electric circuits, require that we retreat from
(\ref{eq.lagr.negle.rest.mass}) back to the Darwin Lagrangian in
the form (\ref{eq.darw.lagr.density.form}), which we had before we
neglected rest mass inertia. From there the equations of motion
for electromechanical systems can be found by adding constraints,
or assumptions, about the motion, thereby reducing the number of
degrees-of-freedom, in the way familiar from analytical mechanics.

If macroscopic matter moves one must, of course, add the
Lagrangian corresponding to that motion. Further, the induction
and capacitance coefficients may now depend on the mechanical
degrees-of-freedom corresponding to the motion of thin wires and
extended conductors of the system, since this changes its
geometry. For an energy conserving system one then typically
arrives at a Lagrangian of the form,
\begin{equation}\label{eq.lagr.mech.plus.lin.electric.circuit}
\lagr(q,e,\dot q, \dot e) = T(q, \dot q) - V(q) +  \half
\sum_{k=1}^n \sum_{l=1}^n \left[ L_{kl}(q) \dot e_k \dot e_l -
C^{-1}_{kl}(q) e_k e_l \right],
\end{equation}
and this will be general enough for our purposes. One notes that
if charged conductors move this produces magnetic effects which
may have to be handled. In many cases, however, the speed of this
motion will be such that the magnetic effect is negligible. In
more general cases one may, of course, also have coupling terms
between $\dot q$ and $\dot e$. In case of doubt the safe method is
to start with the Darwin Lagrangian (\ref{eq.LtotNoRad2}) and
introduce relevant constraints and idealizations. A couple of
examples of this procedure can be found in Ess{\'e}n
\cite{essen05,essen06}.

Electromechanical systems are treated \eg\ in the books by
Ne\u{i}mark and Fufaev \cite{BKneimark}, Wells \cite{BKwells}, and
Gossick \cite{BKgossick}.  Articles discussing various aspects of
these systems are
\cite{wells,ogar&dazzo,hadwich&pfeiffer,basic&al}. We now proceed
to some concrete examples of magnetomechanical systems.

\subsection{Extension of current carrying ring}
When a current $i$ flows in a conducting circular loop, or ring,
its radius will increase somewhat. This is due to the reaction
forces to the forces needed to bend the current that, due to
inductive inertia, otherwise would move in a straight line. Let us
calculate this increase in radius.
\begin{figure}[h]
\centering
\includegraphics[width=230pt]{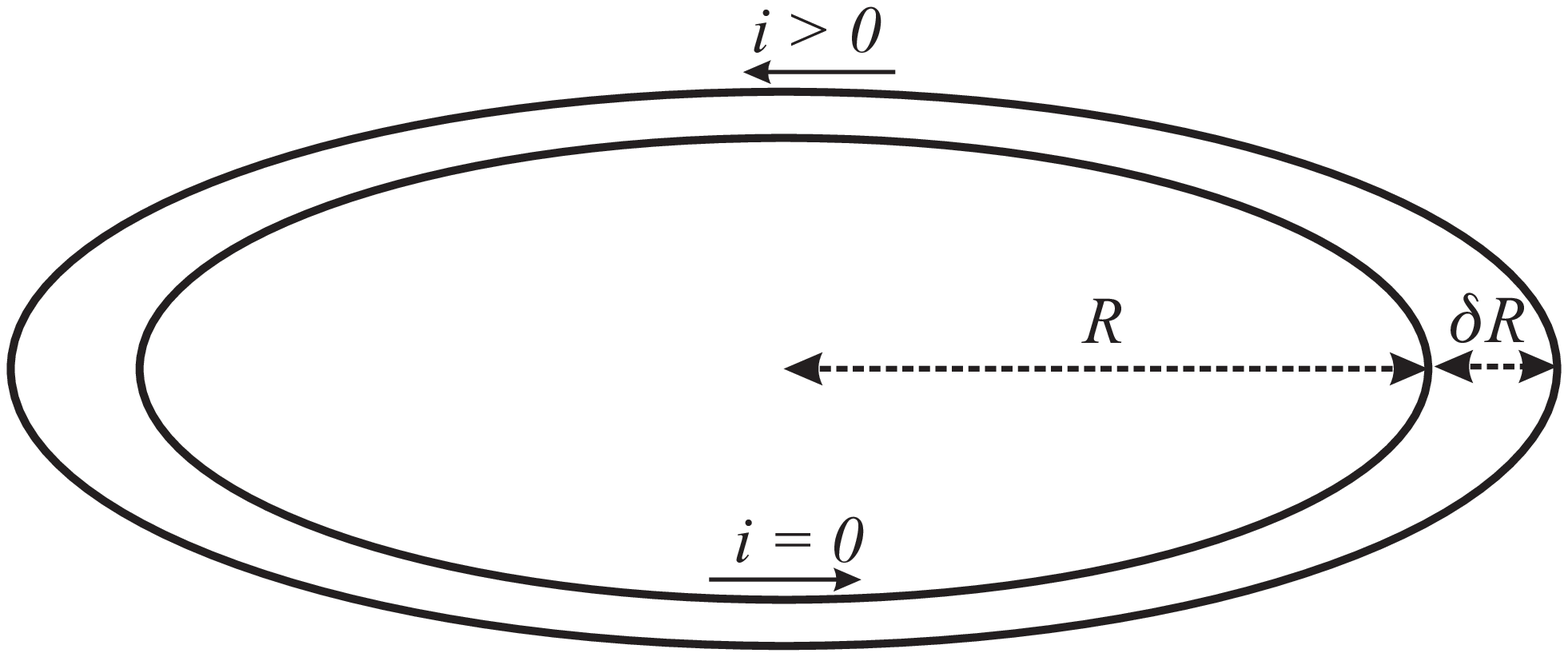}
\caption{\small Notation for the ring extension problem. When the
current  $i =\dot e$ flowing in the ring is zero its radius is
$R$. When the current is increased the radius becomes $R+ \delta R
= R(1+\xi)$. \label{fig.RingExtension}}
\end{figure}

The self-inductance of a ring made of thin wire of circular
cross-section is, see Appendix \ref{app.3} below,
\begin{equation}\label{eq.self.induct.ring}
L_{\rm c}(R)=4\pi R\, [\, \ln(8R/\lambda) - (7/4)]/c^2,
\end{equation}
where $R$ is the ring radius (at zero current), and $\lambda$
is the  radius of the thin wire. Assume that the wire ring can
be treated as an elastic with stiffness $k$. The energy
required to increase its length, $\ell=2\pi R$, by $\delta\ell
= 2\pi\, \delta R$ is then,
\begin{equation}\label{eq.pot.elast.pot.enrgy.ring}
V(\delta R) = \half k (\delta\ell)^2 = \half k (2\pi)^2 (\delta R)^2 .
\end{equation}
If we introduce the notation, $\kappa=4 \ln(8R/\lambda)$, and,
$\xi = \delta R/R$, for the relative change in radius, we get the
self-inductance,
\begin{equation}\label{eq.self.induct.ring1}
L_{\rm c}(\xi)=\pi R(1+\xi)\, [(\kappa-7) + 4 \ln(1+\xi)]/c^2,
\end{equation}
for a ring as a function of the relative extension $\xi$ of the
radius. The elastic potential energy is,
\begin{equation}\label{eq.pot.elast.pot.enrgy.ring1}
V(\xi) = \half k (\delta\ell)^2 = \half k\, \ell^2 \xi^2 ,
\end{equation}
where $k$ is the stiffness, or spring constant. We can now study
the this  two degree-of-freedom magnetomechanical system, the
degrees-of-freedom being $e$, with $i=\dot e$, and $\xi$.

The Lagrangian will have the form given in Eq.\
(\ref{eq.lagr.mech.plus.lin.electric.circuit}) and becomes,
\begin{equation}\label{eq.lagr.ring.extension}
\lagr(\xi,\dot \xi, \dot e) = \half m R^2 \dot\xi^2 - V(\xi) +
\half L_{\rm c}(\xi)\dot e^2,
\end{equation}
where the first term is the kinetic energy of the ring, of mass
$m$, due to time dependent radius. The corresponding Hamiltonian,
including the kinetic energy of radial ring oscillations, the
elastic energy, and the magnetic energy, is thus,
\begin{equation}\label{eq.hamilton.ring.ext}
\hami(\xi,p_{\xi},p) =  \frac{p_{\xi}^2}{2mR^2} + \half k \,\ell^2
\xi^2 + \frac{p^2}{2 L_{\rm c}(\xi)}.
\end{equation}
Here $p=L_{\rm c} \dot e$ is the magnetic flux, divided by $c$,
through the  ring due to the current $\dot e=i$. The generalized
momentum, $p_{\xi} = m R^2 \dot \xi$, is the momentum of
corresponding to radial motion of the ring.

Assume that we increase the flux, $p$, through the ring slowly
without  exciting radial oscillations. We might think of it as
lying on a horizontal lubricated surface that dissipates kinetic
energy keeping $p_{\xi}=0$. Let us calculate the ring extension.
It will correspond to the minimum of the sum of the elastic and
the magnetic energies ($\ell = 2\pi R$),
\begin{figure}[h]
\centering
\includegraphics[width=200pt]{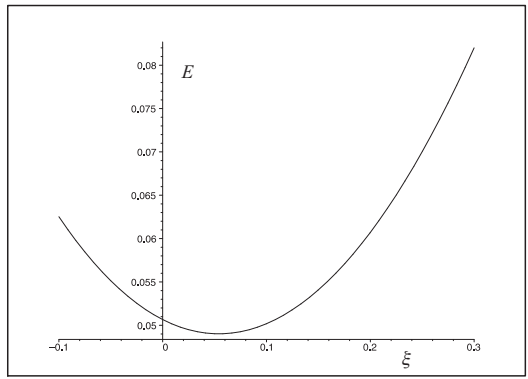}
\caption{\small The function $E(\xi,p)$ of equation
(\ref{eq.energy.ring.mag.plus.elast}) plotted for $R=100\lambda$,
so that $\kappa\approx 27$, and with, $k=\ell=p=c=1$, as a
function of $\xi$. \label{fig.RingExtensionXi}}
\end{figure}
\begin{equation}\label{eq.energy.ring.mag.plus.elast}
E(\xi, p) = \frac{k}{2} (\ell \xi)^2 + \frac{p^2 c^2}{\ell
(1+\xi)\, [(\kappa-7) +  4 \ln(1+\xi)]} ,
\end{equation}
since this minimum corresponds to the minimum of the effective
potential  for the $\xi$-motion, $p$ being a constant of the
motion. A graph of this function is shown in Fig.\
\ref{fig.RingExtensionXi}.

We first differentiate $E(\xi, p)$ with respect to $\xi$. To find
the  minimum we then wish to solve, $\dfd E/\dfd \xi =0$, for
$\xi$, but this equation does not give any simple analytic root.
We therefore first expand in the presumably small parameter $\xi$
and keep the constant and the linear term. The resulting equation
is trivial to solve. Some algebraic rewriting make it possible to
write the root in the form,
\begin{equation}\label{eq.root.ring.ext.xi}
\xi_0(p) = \frac{(\kappa-3)(\kappa-7)c^2 p^2}{(\kappa-7)^3  \ell^3
k+2[7+ (\kappa-4)^2] c^2 p^2} ,
\end{equation}
where $\kappa=4 \ln(8R/\lambda)$. This is thus, to first order,
the relative extension, $\xi=\delta R/R$, of a conducting ring of
radius $R$, cross-sectional radius $\lambda$, and elastic constant
(stiffness) $k$, through which the current, $i=p/L_{\rm c}$,
flows. The ring extension problem is also treated in Landau and
Lifshitz, vol. 8 \cite{BKlandau8}, but in a more complicated way.

\subsection{Parallel coaxial circular current loops}
The mutual inductance of two parallel coaxial rings, of radius
$R_1$ and  $R_2$, a distance $z$ apart, is given by (Becker
\cite{BKbecker}),
\begin{equation}\label{eq.mut.induct.paral.adj.rings}
L_{12} = \frac{2\pi }{c^2}  \int_{0}^{2\pi}   \frac{R_1 R_2
\cos\varphi\, \dfd \varphi }{\sqrt{ z^2+R_1^2+R_2^2-2R_1 R_2
\cos\varphi}} .
\end{equation}
\begin{figure}[h]
\centering
\includegraphics[width=230pt]{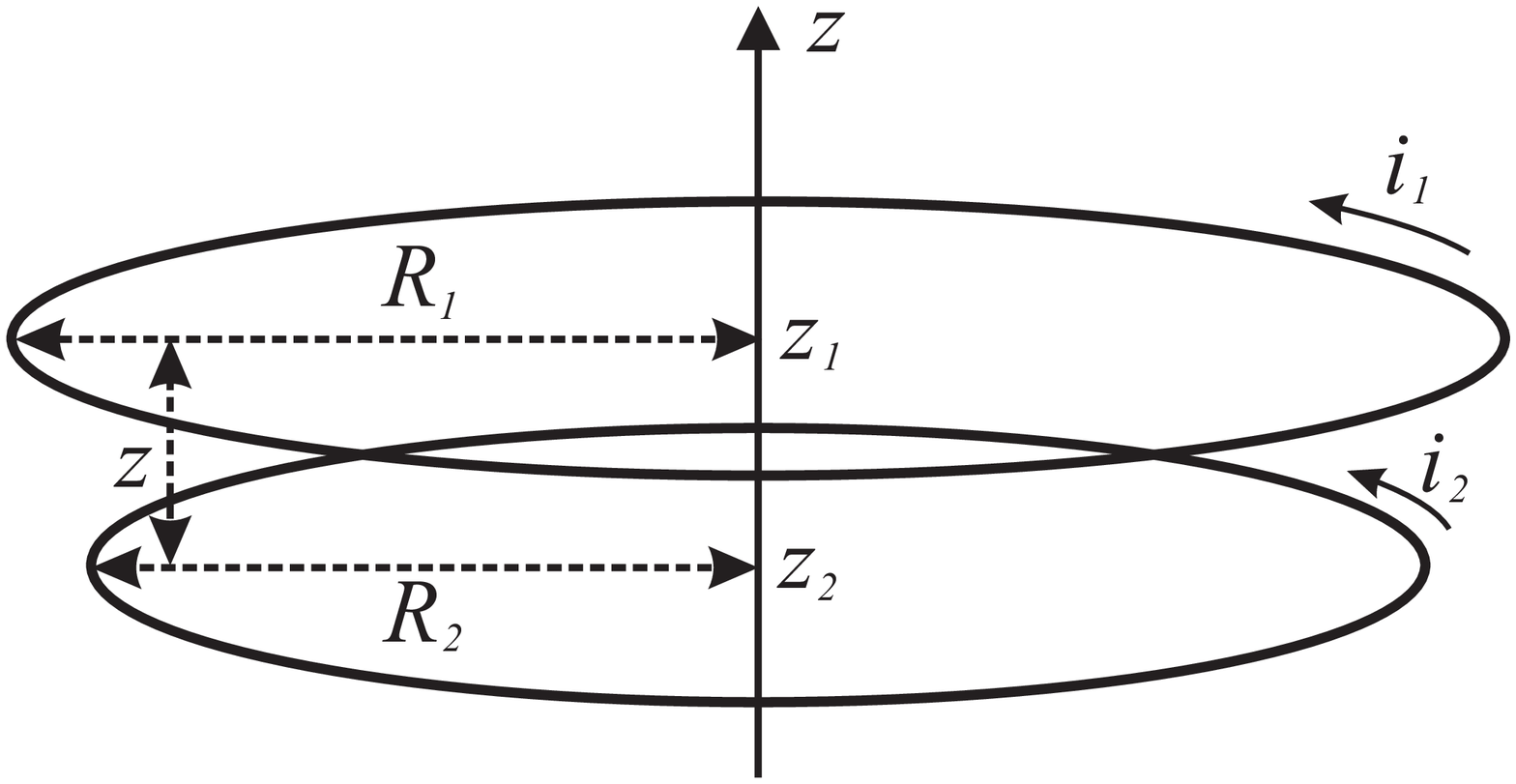}
\caption{\small Notation for the two ring interaction problem.
It is assumed that they can slide freely along the $z$-axis
remaining parallel and coaxial. They have radii $R_1$ and $R_2$ and
the distance between their planes is $z =z_1 - z_2$. The currents in them are
$i_k = \dot e_k,\; k=1,2.$
\label{fig.TwoRingsNear}}
\end{figure}
The integral can be evaluated exactly in terms the complete elliptic
integrals,
\begin{eqnarray}
\label{eq.def.complete.ellipt.int.K}
{\rm K}(k) &=& \int_{0}^{\pi/2} \frac{\dfd\theta}{\sqrt{1 -k^2
\sin^2\!\theta}}, \\
\label{eq.def.complete.ellipt.int.E} {\rm E}(k) &=&
\int_{0}^{\pi/2}\sqrt{1 -k^2 \sin^2\!\theta} \,\dfd\theta .
\end{eqnarray}
Putting,
\begin{equation}\label{eq.eta.def.z.R}
k =\sqrt{ \frac{4 R_1 R_2}{(R_1+R_2)^2+z^2} }
\end{equation}
one obtains,
\begin{equation}\label{eq.mut.induct.paral.equal.adj.rings1}
L_{12} = \frac{4\pi}{c^2}\sqrt{R_1 R_2} \left\{ \frac{2}{k}
\left[{\rm K}(k) - {\rm E}(k)\right]-k {\rm K}(k)  \right\} ,
\end{equation}
for the mutual inductance.

Assuming that the rings can slide along the $z$-axis we now get
the Lagrangian $\lagr=\lagr(z_1, z_2, \dot z_1, \dot z_2, \dot
e_1, \dot e_2)$ of this four degree-of-freedom system in the form,
\begin{equation}\label{eq.two.rings.lagr}
\lagr = \half m_1 \dot z_1^2 + \half m_2 \dot z_2^2 + \half L_{\rm
c1} \dot e_1^2 + \half L_{\rm c2} \dot e_2^2 + L_{12}(z_1-z_2)
\dot e_1 \dot e_2 .
\end{equation}
We first do the well known transformation to center of mass and
relative coordinates: $Z= (m_1 z_1 + m_2 z_2)/M, z= z_1 - z_2$,
where $M=m_1 + m_2$. The Lagrangian $\lagr =\lagr( z, \dot z, \dot
Z, \dot e_1, \dot e_2)$ is now,
\begin{equation}\label{eq.two.rings.lagr.rel.coord}
\lagr = \half M \dot Z^2 + \half m \dot z^2 + \half L_{\rm c1}
\dot e_1^2 + \half L_{\rm c2} \dot e_2^2 + L_{12}(z) \dot e_1 \dot
e_2 ,
\end{equation}
where $m=m_1 m_2/M$ is the reduced mass. We now study this system.

\subsection{Force per length between parallel constant currents}
\label{attrac.parall.curr}First, assume that, we maintain constant
current in both rings: $\dot e_k =I_k, \; k=1,2$. We note that
this is a holonomic (integrable) constraint since it can be
integrated to give, $e_k = e_k(0)+ I_k t$, for the generalized
coordinates. It is thus holonomic, but not time-independent (\ie\
not scleronomic in traditional terminology). Since we will mostly
introduce this constraint here for cyclic (ignorable) coordinates,
\ie\ coordinates that do not appear explicitly in the Lagrangian,
the time dependence of the constraint will not be manifest in the
appearance of Lagrangian. This is a peculiarity of the type of
system treated here, which thus may formally appear conservative,
even though it is not physically conservative. External energy is
normally needed to maintain constant current, even for ideal
conductors.

With this constraint there are only two degrees-of-freedom, $Z$
and $z$, and the Lagrangian (\ref{eq.two.rings.lagr.rel.coord})
gives,
\begin{equation}\label{eq.two.rings.lagr.curr.fixed}
\lagr = \half M \dot Z^2 + \half m \dot z^2 + I_1 I_2 L_{12}(z),
\end{equation}
where we have discarded the two constants due to the self
inductances. This is now a simple system in which the center of
mass $Z$-motion is trivial, and where $V_I (z) = -I_1 I_2
L_{12}(z) $ acts as potential energy of the $z$-motion. Assuming
that $R_1=R_2=R$ and that $z \ll R$ we find from expansion that,
\begin{equation}\label{eq.L12.equal.large.rings}
 L_{12}(z) \approx \frac{4\pi R}{c^2} \left[
\ln\left(\frac{8R}{z}\right) -2 \right] = -  \frac{4\pi R}{c^2}
\ln z +\, \mbox{constant}.
\end{equation}
We throw away the constant term and find, in this approximation,
the potential for the relative motion,
\begin{equation}\label{eq.pot.rel.mot.large.rings}
V_I(z) = \frac{4\pi R}{c^2} I_1 I_2 \ln z.
\end{equation}
This means that,
\begin{equation}\label{eq.force.per.length.betw.rings}
\frac{F}{\ell}  = -\frac{1}{2\pi R}\frac{\dfd V_I}{\dfd z} = -
\frac{2}{c^2} \frac{I_1 I_2}{z}
\end{equation}
is the force per unit length between the rings (of length $\ell =
2\pi R$) assuming $z \ll R$. One recognizes this as the standard
expression for the force per length between parallel currents. The
minus sign means that it is attractive when $I_1 I_2 >0$, \ie\ for
parallel currents, otherwise repulsive. This force between
parallel current carrying wires has been much discussed in the
pedagogical literature
\cite{webster,folmsbee&moran,gabuzda1,gabuzda2,amrani,kampen} but
the analytical mechanical approach presented here does not seem to
have received much attention. The importance of the problem
originates in the fact that the definition of the ampere, the SI
unit of electric current, is based on this type of force
measurement.

\subsection{Relative oscillation of two rings of current}
Assuming constant currents is not natural in this type of problems
where we assume energy conservation. In general maintaining
constant current requires that work is done by an external \emf.
For two perfectly conducting rings of modest size it is more
natural to assume an isolated system of constant energy. We return
to the Lagrangian (\ref{eq.two.rings.lagr.rel.coord}) and note
that since the coordinates (charges) $e_1$ and $e_2$ do not
appear, the corresponding generalized momenta, $p_1$ and $p_2$,
given by,
\begin{equation}
\frac{\partial \lagr}{\partial \dot e_1}\equiv p_1 = L_{\rm c1}\,
\dot e_1 + L_{12}(z)\, \dot e_2 , \;\;\;  \frac{\partial
\lagr}{\partial \dot e_2}\equiv p_2 = L_{\rm c2}\, \dot e_2 +
L_{12}(z)\, \dot e_1 ,
\end{equation}
are conserved. Solving these for the currents,
\begin{equation}
\dot e_1 =
 \frac{ L_{\rm c2}\, p_1 -L_{12}(z)\,
 p_2 }{  L_{\rm c1} L_{\rm c2}
 - L^2_{12}(z) } , \;\;\;   \dot e_2 =
 \frac{ L_{\rm c1}\, p_2 -L_{12}(z)\, p_1 }{  L_{\rm c1} L_{\rm c2}
 - L^2_{12}(z) } ,
\end{equation}
we can proceed to find the Hamiltonian corresponding to the
Lagrangian (\ref{eq.two.rings.lagr.rel.coord}). This gives,
\begin{equation}\label{eq.ham.of.two.rings}
\hami = \frac{p_Z^2}{2M} + \frac{p_z^2}{2m} + \frac{ L_{\rm c2}
p_1^2 + L_{\rm c1} p_2^2 -2 L_{12}(z) p_1 p_2 }{2 [L_{\rm c1}
L_{\rm c2}
 - L^2_{12}(z)] }.
\end{equation}
We now wish to compare the interaction of the two rings for the
case of constant currents $\dot e_1 = \dot e_2 = I$, and for the
case of constant momenta,
\begin{equation}\label{eq.values.of.momenta.two.rings}
p_1 = [L_{\rm c1} + L_{12}(0)]  I, \;\;\; p_2 = [L_{\rm c2} +
L_{12}(0)] I,
\end{equation}
assuming currents $I$ at $z=0$.

\begin{figure}[h]
\centering {\includegraphics[width=230pt]{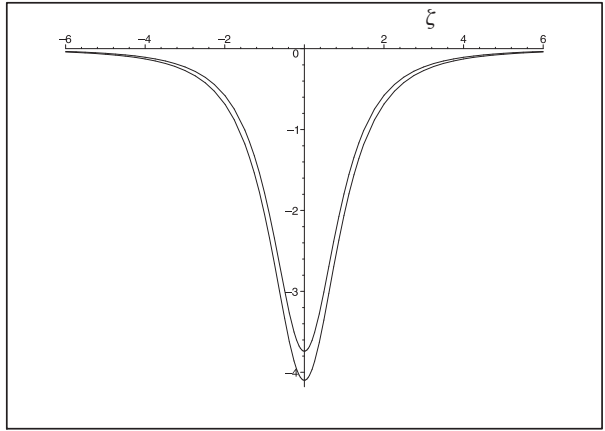}}
\caption{\small Plots of interaction potentials of two rings of
current, see Fig.\ \protect\ref{fig.TwoRingsNear}. For constant
unit currents the potential is $V_I = - L_{12}(\zeta)$ and this is
the upper curve. For constant generalized momenta, assuming unit
currents at $z=\zeta=0$, the lower curve, corresponding to $V_p$
of Eq.\ (\ref{eq.eff.pot.ham.tow.rings}), is obtained.
\label{fig.PotOfTwoRings}}
\end{figure}
For constant currents the interaction potential is simply the
negative of the last term of (\ref{eq.two.rings.lagr.rel.coord}),
\begin{equation}\label{eq.eff.pot.lagr.rings}
V_I(z) = - I^2 L_{12}(z).
\end{equation}
For the Hamiltonian (\ref{eq.ham.of.two.rings}) the potential that
goes to zero at infinity is obtained by subtracting the constant,
$\frac{p_1^2 }{2 L_{\rm c1} } + \frac{p_2^2 }{2 L_{\rm c2} }$,
from the last term. This gives the potential,
\begin{equation}\label{eq.eff.pot.ham.tow.rings}
V_p(z) =V_I(z) \frac{   \left(1 + \frac{L_{12}(0)}{ L_{\rm c1}}
\right)\left(1 + \frac{L_{12}(0)}{ L_{\rm c2}} \right) -
\frac{L_{12}(z)}{2}  \left[ \frac{\left(1 + \frac{L_{12}(0)}{
L_{\rm c2}} \right)^2}{L_{\rm c1}}  + \frac{\left(1 +
\frac{L_{12}(0) }{ L_{\rm c1} } \right)^2}{L_{\rm c2}}
 \right]  }{ 1 - \frac{L^2_{12}(z)}{ L_{\rm c1}L_{\rm c2} } },
\end{equation}
for relative $z$-motion of the closed conservative system of two
perfectly conducting rings.

To get definite results and compare the two expressions we now
introduce specific values of the parameters. We use
(\ref{eq.self.induct.ring}) for the self-inductances of the rings,
$L_{\rm cj}, \; j=1,2$, and expressions (\ref{eq.eta.def.z.R}) and
(\ref{eq.mut.induct.paral.equal.adj.rings1}) for the mutual
inductance $L_{12}(z)$ of the two rings. For definiteness we put,
$R_1 =\ell$, $R_2 =(9/10) \ell$, $\lambda=\ell/20$, and use the
dimensionless distance, $\zeta = z/\ell$, instead of the distance
$z$ between the planes of the rings, see Fig.\
\ref{fig.TwoRingsNear}. With these choices, and putting
$I=\ell=c=1$, we get the interaction potentials shown in Fig.\
\ref{fig.PotOfTwoRings}. The difference is fairly small due two
the fact that the self-inductances are an order of magnitude
larger than the maximum value of the mutual inductance even though
the parameters have been chosen to maximize the difference. A
system for which the difference between a (time dependent)
constant current constraint and a closed conservative system is,
qualitatively and quantitatively, of importance is described in
the following section.

\subsection{Rectangular circuit and the rail gun} Here we study a
system which can be thought of as an idealized rail gun. It
reinforces the lesson in the ring extension example above in
showing that closed loops of current tend to expand. In the
example of the ring extension this can be understood as due to the
inertia of the current. The current tries to go straight but has
to follow the conducting wire and thus there must be a reaction
force from the current on the wire that tends to straighten it
out. Here we will see that this also happens when there are right
angled corners in the circuit.

The self-inductance of a rectangular circuit with side lengths $a$
and $b$ made of flat conducting strips of width $\lambda$ in the
plane of the rectangle has been calculated by Bueno and Assis
\cite{BKbueno&assis,bueno&assis}. Their result is,
\begin{eqnarray}\nonumber
L_{\rm r}(a,b,\lambda) \approx \frac{1}{c^2}\left\{ 4\left[
a\,\ln\left(\frac{2a}{\lambda}\right) +
b\,\ln\left(\frac{2b}{\lambda}\right) \right]
 \right.\\
\label{eq.self.induct.rectangle} \\
\nonumber \left. -4 \left[ a\arcsinh\left(\frac{a}{b}\right) +b
\arcsinh\left(\frac{b}{a}\right) \right] + 8 \sqrt{a^2+b^2}-2(a+b)
\right\},
\end{eqnarray}
where the neglected terms are of order $\lambda \left\{ {\cal
O}[(\lambda/a)^2] + {\cal O}[(\lambda/b)^2] \right\}$. We now
introduce $a=x=b\xi$ and $\lambda=b/10$ and assume that $\xi>1$
so that we can expand in $1/\xi$. This gives,
\begin{equation}\label{eq.exp.induct.rect.x}
L_{\rm r}(\xi) \approx \frac{b}{c^2} \left( [4 \ln(10)+6]\,\xi + [4
\ln(20)-2] -\frac{1}{\xi} + \frac{1}{24 \xi^3} \right),
\end{equation}
\begin{figure}[h]
\centering {\includegraphics[width=250pt]{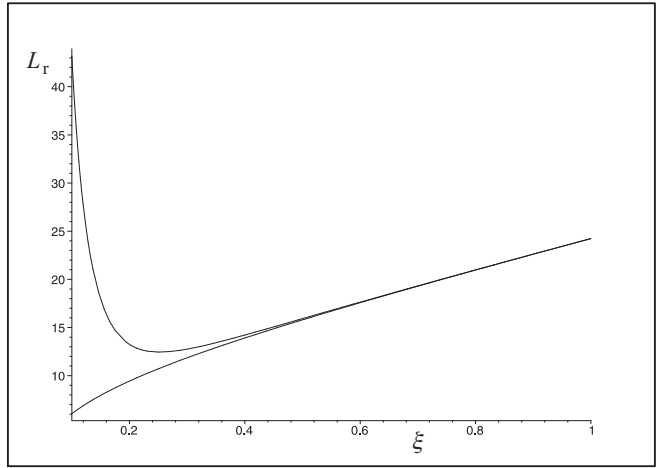}}
\caption{\small Comparison of the two approximations of the
self-inductance $L_{\rm r}(\xi)$ of a rectangle. Here $b=c=1$ and
the lower curve is the approximation in
(\ref{eq.self.induct.rectangle}). The upper curve is the truncated
series approximation in  (\ref{eq.exp.induct.rect.x}). The
approximation is excellent for $0.5 < \xi$.
\label{fig.Rectangel.self.ind}}
\end{figure}
where terms of order $1/\xi^5$ and higher have been neglected. The
two expressions (\ref{eq.self.induct.rectangle}) and
(\ref{eq.exp.induct.rect.x}) are compared in the plot of Fig.\
\ref{fig.Rectangel.self.ind}.

\begin{figure}[h]
\centering
\includegraphics[width=250pt]{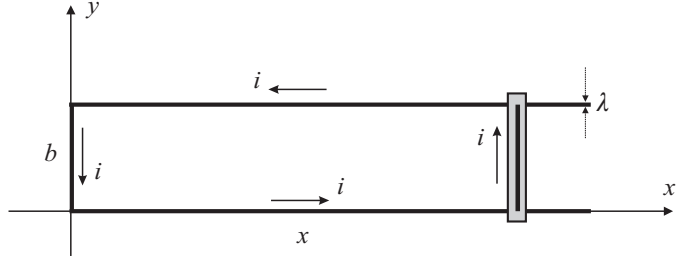}
\caption{\small The self-inductance $L_{\rm r}$ of this
rectangular current loop is given by Eq.\
(\ref{eq.self.induct.rectangle}) with $x=a$. A simple model of a
rail gun is obtained when the right hand edge of the rectangle can
slide, with negligible friction, in the direction of the $x$-axis.
\label{fig.RailGun}}
\end{figure}

The Lagrangian of the two degree-of-freedom system shown in Fig.\
\ref{fig.RailGun} is then,
\begin{equation}\label{eq.rail.gun.lagr}
\lagr = \half m  b^2 \dot \xi^2 + \half L_{\rm r} (\xi) \dot e^2 ,
\end{equation}
where, $\dot e = i$, is the current in the circuit. The coordinate
$e$ is absent (cyclic) so the generalized momentum,
\begin{equation}\label{eq.gen.mom.cons.railg}
p \equiv \frac{\partial \lagr}{\partial \dot e} = L_{\rm r}(\xi)
\dot e ,
\end{equation}
is conserved, assuming perfectly conducting parts. The conserved
Hamiltonian becomes,
\begin{equation}\label{eq.rail.gun.ham}
\hami(\xi,p_\xi,p) = \half \frac{p_\xi^2}{m  b^2 } + \half
\frac{p^2}{L_{\rm r} (\xi)} = \half \frac{p_\xi^2}{m  b^2 } + V_p(\xi),
\end{equation}
where $p_\xi=mb^2 \dot \xi$. Assume that $\dot \xi(0) = 0$ at
$\xi(0) =\xi_0$, so that the conserved energy, $\hami=E$, is $E=
\frac{p^2}{ 2 {L_{\rm r} (\xi_0) }}$. One then finds that,
\begin{equation}\label{eq.speed.of.bar.of.x}
v(\xi) = b\dot \xi(\xi) = p \sqrt{\frac{1}{m}\left(
\frac{1}{L_{\rm r} (\xi_0)} - \frac{1}{L_{\rm r} (\xi)}  \right)},
\end{equation}
is the velocity of the moving bar as function of $\xi = x/b$.
Since the self-inductance goes to infinity with $\xi$ the limiting
velocity of the bar will be $v(\infty) = p/\sqrt{mL_{\rm
r}(\xi_0)}$. The speed of the rail gun projectile is thus
proportional to the conserved magnetic flux $\Phi =c p$ and
inversely proportional to the square root of the self-inductance
of the initial rectangle $L_{\rm r} (\xi_0)$.

Assume instead that we maintain a constant current, \ie\ introduce
the constraint $\dot e=I$, in the rectangular circuit. A look at
the Lagrangian (\ref{eq.rail.gun.lagr}) then shows that the there
is, formally, a conserved energy,
\begin{equation}\label{eq.cons.energy.cons.curr.rail.gun}
{\cal E}(\xi, \dot\xi) = \half m  b^2 \dot \xi^2 - \half I^2
L_{\rm r} (\xi).
\end{equation}
The bar will move in the potential $V_I(\xi) = - \half I^2 L_{\rm
r}(\xi)$ and thus accelerate indefinitely (until the end of the
rails). Though this still is formally a conservative system it is
clear that energy must be continuously feed into the system to
achieve the continuous acceleration of the bar. In fact charge in
the system increases linearly, $e(t) = e(0) + It$, according to
the implied time dependent constraint.

The treatment of the rail gun above is essentially new as far as
the author knows, but this type of system has certainly been
discussed both in the pedagogical and technical literature (see
\eg\ Knoepfel \cite{BKknoepfel}). Some examples from pedagogical
journals are \cite{namias,robson&sethian,jonesr}. Because of the
many potential applications of rail guns there is a huge technical
literature on the subject. A technical treatment using the
Lagrangian formalism of electromechanical systems is by Hively and
Condit \cite{hively&condit}.

\section{On the nature of magnetic energy} \label{conclus}
It should be clear from the above examples that the methods of
analytical mechanics can be quite useful in treating
electromechanical, and in particular magnetomechanical, systems.
There is no need to first find the fields and then the forces from
these. Instead both steps are integrated into a single formalism.

While a Lagrangian with no explicit time dependence corresponds to
a conservative system, one should note that a constant current
constraint may only be formally energy conserving. External work
may be needed to keep current constant. Consider the magnetic
(inductive) part of the energy expression
(\ref{eq.energ.lin.electric.circuit}),
\begin{equation}\label{eq.energ.lin.electric.circuit.mag}
{\cal E}_{\rm L}(q,\dot e)= \half \sum_{k=1}^n \sum_{l=1}^n L_{kl}
(q) \dot e_k \dot e_l ,
\end{equation}
where $q$ represents mechanical degrees-of-freedom. Should the
currents be kept constant, $\dot e_k = I_k =$ constant, we find
that this term becomes the negative of an effective potential for
the $q$-motion,
\begin{equation}\label{eq.energy.mag.eff.pot}
V_I(q)=- \half \sum_{k=1}^n \sum_{l=1}^n L_{kl} (q) I_k I_l  =
-{\cal E}_{\rm L}(q,I),
\end{equation}
in a full Lagrangian of the form $\lagr_I (q, \dot q) = T(\dot q)
- V_I(q)$. For this Lagrangian an equilibrium position corresponds
to a minimum of $V_I(q)$. Evidently this corresponds to a maximum
of the energy (\ref{eq.energ.lin.electric.circuit.mag}). We have
thus arrived at the result that in electromechanical systems, for
which current is kept constant, magnetic energy will tend to a
maximum, when the system tends to its equilibrium.

The above result does not seem well known, even though explicitly
stated in the textbook by Greiner \cite{BKgreinerCED}. It is also
in accord with Woltjer's \cite{woltjer} assumption that self
organized states of a plasma should correspond to maxima of
magnetic energy. Mehra and De Luca \cite{mehra}, on the other
hand, made computer simulations of a plasma minimizing the
velocity space form of the Darwin energy
(\ref{eq.LtotNoRad2.energy}). They then found surprising results
indicating that anti-parallel currents attracted each other. From
our example above we know that it is the other way around. In
conclusion, the velocity space form of the magnetic energy tends
to become maximized.

Why would magnetic energy be so different from other forms of
energy  which usually tend to minima in equilibrium states? We can
resolve this conundrum by recalling that there is also the
Hamiltonian form of magnetic energy. The Hamiltonian corresponding to the Darwin energy is discussed in \cite{essen&nordmark}. For circuits we have,
\begin{equation}\label{eq.hamiltonian.LC.circuit.mag}
\hami_{\rm L}(q,p) = \half \sum_{k=1}^n \sum_{l=1}^n
L^{-1}_{kl}(q) p_k p_l .
\end{equation}
We learned that the generalized momenta $p_k$, proportional to
magnetic fluxes, are conserved if the corresponding charge $e_k$
is cyclic. For constant momenta $p_k =p^0_k$ the effective
potential for the $q$-motion is,
\begin{equation}\label{eq.hamiltonian.LC.circuit.mag.pot}
V_p(q) = \hami_{\rm L}(q,p^0) .
\end{equation}
This form of the magnetic energy is minimized when the $q$-motion
maximizes the inductive coefficients. The Hamiltonian and the
Lagrangian formalisms thus correspond to two different energy
concepts. In the case of magnetic energy, which is a kinetic
energy, the coefficients of the generalized velocities, and the
coefficients of the generalized momenta are each others inverse,
which means that the minimum of one is the maximum of the other.

Schwinger \etal\ \cite{BKschwinger&al} is the only text that
discusses the above facts concerning magnetic energy briefly. They
also point out that the velocity space form of energy seems to
give the erroneous idea that parallel currents repel. Being clear
about which variables are held constant when one searches for a
minimum is essential. Stating that magnetic energy is a maximum or
a minimum for some configuration, as has been done
\cite{fiolhais}, is meaningless unless this is made explicit.

\appendix
\section{Appendices}
Here the generalized capacitance coefficients and the inductance
coefficients of linear circuit theory are derived and explained.
Several results here are new. Even if the existence of these
coefficients is stated in many texts, the discussion of their
properties and physical origin usually is rather brief.

\subsection{Energy of a system of charged conductors and
generalized capacitance coefficients}
Consider the electric energy expressed in the form,
\begin{equation}\label{eq.el.stat.energy}
W_e =\half \int  \varrho\phi \, \dfd V .
\end{equation}
The potential, $\phi$, is a solution of Poisson's equation,
\begin{equation}\label{eq.poisson}
\nabla^2 \phi = -4\pi \varrho,
\end{equation}
where it is assumed that the solution must go to zero at large
distance from the region where the charge density, $\varrho$,
is located. For point particles the solution to this equation
is well known. Assuming that, $\varrho=\varrho_i(\vecr) = e_i
\delta(\vecr-\vecr_i)$, the solution is, $\phi=
\phi_i(\vecr)=e_i/|\vecr-\vecr_i|$. For $N$ particles the
charge density is the sum, $\varrho = \sum_i^N \varrho_i$, and
the solution is simply a superposition of such solutions,
$\phi(\vecr) = \sum_i^N \phi_i$, due to the linearity of the
Poisson's equation. When these results are inserted into
(\ref{eq.el.stat.energy}) one obtains the result,
\begin{equation}\label{eq.el.stat.energy.point.part}
W_e =\half \sum_{i,j=1}^N \frac{e_i e_j }{r_{ij} },
\end{equation}
where, $r_{ij}=|\vecr_j - \vecr_i|$, are the distances between
the particles, and where it is necessary to exclude the case,
$i=j$, to get a finite result.

Assume now that we, instead of particles, have a set of $n_c$
fixed conductors occupying the (compact) volumes $V_k, \;
(k=1,\ldots,n_c)$ and ask: what is the potential $\phi$ and the
energy $W_e$ of the system if we put charges $e_k$ on all, or
some, of these (isolated) conductors? We know that the charge on
each conductor must be distributed on its surface in such a way
that the electric potential, $\phi(\vecr)$, is constant on the
surface, and the interior, of  each conductor $V_k$. Otherwise
current will flow until it becomes constant, since a gradient
implies presence of electric field.
\begin{figure}[h]
\centering
\includegraphics[width=200pt]{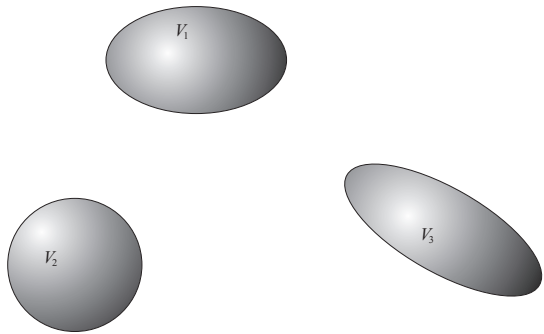}
\caption{\small Here we study the potential arising from a
system of conductors $V_i$ which either have a distribution
$\varrho^*_{ii}$ of unit net charge or are only polarized with
zero net charge redistributed to a density $\varrho^*_{ij}$, so
as to make the potential constant on each conductor. In the
figure $i,j = 1,2,3$. \label{fig.Conductors}}
\end{figure}

For the particle problem the charge density, $\varrho = \sum_i^N
e_i \varrho^*_i$, with, $\varrho_i^* = \delta(\vecr-\vecr_i)$,
gave rise to the potential, $\phi = \sum_i^N e_i \phi_i^*$, where,
$\phi^*_i = 1/|\vecr-\vecr_i|$. We now wish to determine
corresponding densities $\varrho_i^*$ and potentials $\phi_i^*$
for the case of isolated conductors, of given size, shape, and
position. Assume that we place a unit charge on conductor $V_i$
while all the other conductors have zero charge. We first note
that this does not mean that the charge densities on the other
conductors are identically  zero since charge on one of the
conductors will induce a polarizing redistribution of the charge
on the others. The total charge density for this case is thus,
\begin{equation}\label{eq.charge.dens}
\varrho^*_i = \sum_{k=1}^{n_c} \varrho^*_{ik} ,
\end{equation}
where,
\begin{equation}\label{eq.normalization.of.ij.dens}
\int_V \varrho^*_{ik} \dfd V = \delta_{ik} = \left\{
\begin{array}{ccc}  1 & \mbox{ if } & i = k \\ 0 & \mbox{ if } & i
\neq k \end{array} \right.
\end{equation}
\ie\ the net charge is unity on conductor $i$ and zero on the
other  conductors. Our assumption means that these densities must
obey,
\begin{equation}\label{eq.poisson.special}
\nabla^2 \phi^*_i = -4\pi \varrho^*_i,
\end{equation}
with $\phi^*_i(\vecr)$ given by,
\begin{equation}\label{eq.pot.energy.of.set.of.charged.conduct}
\phi^*_i(\vecr) = \int_V \frac{\varrho_i^*(\vecr')\,\dfd
V'}{|\vecr-\vecr'|}  = \sum_{k=1}^{n_c} \int_{V_k}
\frac{\varrho^*_{ik} (\vecr_k)\,\dfd V_k}{|\vecr-\vecr_k|} =
\sum_{k=1}^{n_c} \phi^*_{ik}(\vecr),
\end{equation}
and being such that,
\begin{equation}\label{eq.constancy.phi.star}
\phi^*_i(\vecr_j) =\Gamma_{ij} = \mbox{ const. for all }\;
\vecr_j \in V_j .
\end{equation}
Properly chosen $\varrho^*_{ij}$ must thus produce functions
$\phi^*_i$  are constant on all of the conductors. One notes that
the field $\phi^*_{ii}(\vecr)$ has monopole character, while the
$\phi^*_{ij}(\vecr)$, for $i\neq j$, are of a dipole character.

The general solution to the Poisson equation (\ref{eq.poisson})
for a  system of $n_c$ conductors with charges $e_i$ on the first
$N (\le n_c)$ of them is then,
\begin{equation}\label{eq.gen.solut.syst.conduct}
\phi(\vecr) = \sum_{i=1}^N e_i \phi^*_i(\vecr),
\end{equation}
according to the superposition principle. We see that the
potentials  $\phi^*_{i}$, arising from unit charge on conductor
$i$ and zero on the rest, constitute a basis set of functions for
this problem, analogous to the functions $\phi^*_i =
1/|\vecr-\vecr_i|$ for the point particle problem.

Let us return, now to the energy expression Eq.\
(\ref{eq.el.stat.energy}). Since the charge density $\varrho$ is
non-zero only on the conductors  $V_k$ we find that
\begin{equation}\label{eq.el.stat.energy1}
W_e =\half \int  \varrho\phi \, \dfd V = \half \sum_{k=1}^{n_c}
\int_{V_k} \varrho(\vecr) \phi(\vecr) \,\dfd V .
\end{equation}
Use of (\ref{eq.gen.solut.syst.conduct}) gives us,
\begin{equation}\label{eq.el.stat.energy2}
W_e  = \half \sum_{k=1}^{n_c}  \int_{V_k} \varrho(\vecr)
\sum_{i=1}^N e_i \phi^*_i(\vecr) \,\dfd V,
\end{equation}
but, since $\phi^*_i$ is constant on each of the conductors it can
be  taken out of the integral and replaced by its constant value
on conductor $V_k$,
\begin{equation}\label{eq.el.stat.energy3}
W_e  = \half \sum_{k=1}^{n_c} \sum_{i=1}^N e_i \phi^*_i(\vecr_k)
\int_{V_k} \varrho(\vecr)  \,\dfd V = \half \sum_{j=1}^{N}
\sum_{i=1}^N e_i \phi^*_i(\vecr_j) e_j.
\end{equation}
The second equality here is due to the fact that,
\begin{equation}\label{eq.charge.density.tot.as.sum.star}
\varrho = \sum_{j=1}^N e_j \varrho_j^* =\sum_{j=1}^N e_j
\sum_{l=1}^{n_c} \varrho^*_{jl} ,
\end{equation}
so, according to (\ref{eq.normalization.of.ij.dens}), the integral
of the charge density over $V_k$ is simply $e_k$, for $k=1,\ldots,
N$, and zero for $k=N+1,\ldots, n_c$. So, using
(\ref{eq.constancy.phi.star}), we finally get,
\begin{equation}\label{eq.el.stat.energy4}
W_e   = \half \sum_{i,j=1}^N \Gamma_{ij} e_i e_j ,
\end{equation}
for the energy of the set of conducting bodies. When we compare
this to (\ref{eq.el.stat.energy.point.part}) we see that  the
quantity $\Gamma_{ij}$ represents a kind of effective inverse
distances between the charges on the conductors. As defined in
(\ref{eq.pot.energy.of.set.of.charged.conduct} -
\ref{eq.constancy.phi.star}) this quantity is the constant value
of the potential on body $j$ when unit net charge is distributed
on body $i$ and zero net charge on all the other $n_c -1$ bodies.
Note that bodies with zero charge contribute to the values of the
$\Gamma_{ij}$.

If we denote the constant value of the potential on conductor  $i$
by $\phi_i = \phi(\vecr_i)$ we find from
(\ref{eq.gen.solut.syst.conduct}) that,
\begin{equation}\label{eq.pot.in.terms.of.charge.sum}
\phi_i = \sum_{j=1}^N e_j \phi^*_j(\vecr_i)  = \sum_{j=1}^N e_j
\Gamma_{ji}, \; \;\; \mbox{ for }\; i=1,\ldots,N,
\end{equation}
where (\ref{eq.constancy.phi.star}) was used to get the last
equality.  These linear equations for the potentials can be solved
for the charges. The result can be written,
\begin{equation}\label{eq.charge.in.terms.of.pots.sum}
e_i =  \sum_{j=1}^N  C_{ij} \phi_j, \; \;\; \mbox{ for }\;
i=1,\ldots,N,
\end{equation}
where the coefficients $C_{ij}$ represent the matrix elements of
the inverse  of the matrix with elements $\Gamma_{ij}$. The
$C_{ij}$ are called generalized capacitance coefficients.

\subsection{Energy of a system of current carrying wires and the
inductance coefficients} This is a subject discussed in many
textbooks. Examples are Landau and Lifshitz, vol. 8
\cite{BKlandau8}, Greiner's Classical Electrodynamics
\cite{BKgreinerCED}, and Johnk \cite{BKjohnk}. A more specialized
text is by Knoepfel \cite{BKknoepfel}.

The magnetic energy is given by,
\begin{equation}\label{eq.mag.energy}
W_m =\frac{1}{2c} \int  \vecj\cdot \vecA \, \dfd V .
\end{equation}
If one assumes that all current is flowing in thin wires one can
replace the infinitesimal vector $\vecj\,\dfd V$ with
$i_k\,\dfd\vecr_k$ where $\dfd\vecr_k$ is a line element along the
curve $C_k$ defined by the $k$th wire (filament). This gives,
\begin{equation}\label{eq.mag.energy.wires.sum1}
W_m =\frac{1}{2c} \sum_k  i_k \int_{C_k}  \vecA \cdot \,  \dfd
\vecr_k .
\end{equation}
One notes immediately that if the curve is a closed curve the
corresponding contribution is gauge independent since,
\begin{equation}\label{eq.mag.energy.closed.wires.gauge.indep}
 \oint_{C_k}  \vecA \cdot \, \dfd \vecr_k =  \oint_{C_k}
(\vecA' + \nabla \chi) \cdot \, \dfd \vecr_k =  \oint_{C_k} \vecA'
\cdot \, \dfd \vecr_k .
\end{equation}
For conducting wires that are not closed but instead go between
conductors that provide capacitance the issue of gauge may need to
be resolved.

When there is gauge independence we can use the expression,
\begin{equation}\label{eq.lor.gauge.vec.pot.from.wires}
\vecA(\vecr) = \frac{1}{c} \int  \frac{\vecj(\vecr') \dfd V'
}{|\vecr-\vecr'|} =\sum_l \frac{i_l}{c} \int_{C_l} \frac{ \dfd
\vecr_l }{|\vecr-\vecr_l|},
\end{equation}
for the vector potential from the currents $i_l$ flowing in wires
along the curves $C_l$. Inserting this into
(\ref{eq.mag.energy.wires.sum}) we find that the magnetic energy
is,
\begin{equation}\label{eq.mag.energy.wires.sum}
W_m =\frac{1}{2c^2} \sum_{kl}  i_k i_l \int_{C_k} \int_{C_l}
\frac{\dfd \vecr_k  \cdot \, \dfd \vecr_l}{|\vecr_k-\vecr_l|} .
\end{equation}
So, if we introduce the wire geometry dependent quantities,
\begin{equation}\label{eq.induc.coeff.gen.two.wires.neumann}
L_{kl} =\frac{1}{c^2} \int_{C_k} \int_{C_l}  \frac{\dfd \vecr_k
\cdot \, \dfd \vecr_l}{|\vecr_k-\vecr_l|},
\end{equation}
we find the magnetic energy in the form,
\begin{equation}\label{eq.mag.energy.Lkl.sum}
W_m =\half \sum_{kl} L_{kl} i_k i_l .
\end{equation}
For fixed positions of the wires the magnetic energy is thus a
quadratic form in the currents with constant coefficients
$L_{kl}$. For $l\neq k$ these are called mutual inductances. When
$l=k$ they are self-inductances.

Apart from the problem of lack of gauge invariance for non-closed
wires one must also deal somehow with the logarithmic divergence
of the self-inductance for a truly filamentary wire. If the wire
is assumed to be a mathematical curve of no thickness the
expression (\ref{eq.induc.coeff.gen.two.wires.neumann}) will
diverge when $k=l$. Should one find a way to handle the divergence
there remains the question of the gauge invariance of the
self-inductances. Let us consider these questions by means of an
example.

We first note that if the Darwin expression
(\ref{eq.darwin.vec.pot.curr.dens.int}) for the vector potential,
is used instead of (\ref{eq.lor.gauge.vec.pot.from.wires}), the
induction coefficients become,
\begin{equation}\label{eq.induc.coeff.gen.two.wires.darwin}
L_{kl} =\frac{1}{2 c^2} \int_{C_k} \int_{C_l}  \frac{\dfd \vecr_k
\cdot \, \dfd \vecr_l + (\dfd \vecr_k \cdot \vece_{kl})(\dfd
\vecr_l \cdot \vece_{kl}) }{|\vecr_k-\vecr_l|},
\end{equation}
\ie\ different from the Neumann form
(\ref{eq.induc.coeff.gen.two.wires.neumann}).

\subsection{Self-inductance of a rotating polygon of charged particles}
\label{app.3}One way of handling the divergence in the filamentary
self-induction is to consider the current as due to many charged
particles travelling in the wire, instead of a continuous
distribution of charge. Assume that the current in a circular wire
is due to $N$ particles of charge $e$ forming a regular polygon
that rotates rigidly. We assume as usual that the material of the
wire is of the opposite charge and cancels the charge of the
particles so that effects of Coulomb interactions are negligible.
\begin{figure}[h]
\centering
\includegraphics[width=170pt]{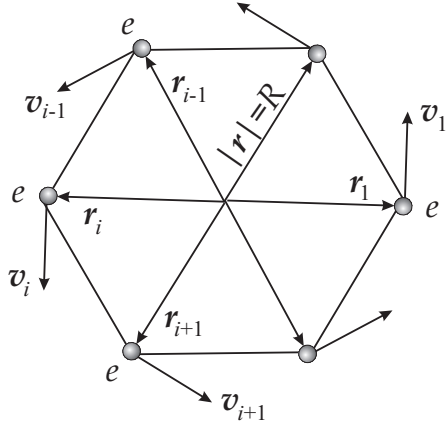}
\caption{\small An example of a polygonal circular current with
$N=6$. \label{fig.PolygonCurrent}}
\end{figure}
Assuming that the circle has radius $R$, positions and velocities
are given by,
\begin{equation}
\label{eq.pos.vel.vectors.meso} \vecr_i(\varphi_i)=R
\vece_{\rho}(\varphi_i), \hskip 0.3cm \mbox{and} \hskip 0.3cm
\vecv_i(\varphi_i,\dot\varphi_i)= R \dot\varphi_i
\vece_{\varphi}(\varphi_i),
\end{equation}
where,  $\vece_{\rho}(\varphi)= \cos\varphi\, \vece_x
+\sin\varphi\,\vece_y$ and $\dot{\vece}_{\rho}
=\dot\varphi\vece_{\varphi}$, as usual. If we further assume that
all the angular velocities $\dot\varphi_i$ are the same, and given
by $\dot\varphi_i=\omega$, and that the angles are $\varphi_i =
(i-1)2\pi/N$ for $i=1,\ldots,N$, we have a rigidly rotating
regular polygon. This is illustrated for $N=6$ in Fig.\
\ref{fig.PolygonCurrent}.

We now calculate the magnetic energy of this system and compare
the expressions obtained using the two different expressions
(\ref{eq.darwin.vec.pot.curr.dens.int}) and
(\ref{eq.lor.gauge.vec.pot.from.wires}) for the vector potential.
Simple calculations show that the magnetic energy can be written,
\begin{equation}\label{eq.mag.energy.polygon}
W_m = \frac{e}{2c} \sum_{i=1}^N v_i A_i ,
\end{equation}
where all $v_i=R\dot\varphi_i = R\omega$, and where,
\begin{equation}\label{eq.vec.pot.polygon}
A_i = \frac{e}{c R} \sum_{j\neq i}^N v_j V_{\varphi}(\varphi_i
-\varphi_j) .
\end{equation}
Here $V_{\varphi}$ is given by,
\begin{equation}
\label{eq.V.phi.darw}
V_{\varphi}^D(\varphi)=\frac{1}{4}\frac{1+3\cos\varphi}{\sqrt{
2(1-\cos\varphi)}},
\end{equation}
in the Darwin case (\ref{eq.darwin.A.ito.velocity}) and by,
\begin{equation}
\label{eq.V.phi.lorenz}
V_{\varphi}^L(\varphi)=\frac{\cos\varphi}{\sqrt{
2(1-\cos\varphi)}},
\end{equation}
in the Lorenz case (\ref{eq.lor.gauge.vec.pot.from.wires}).
Because of the symmetry all terms in the sum
(\ref{eq.mag.energy.polygon}) are equal and we find,
\begin{equation}\label{eq.mag.energy.polygon.simpl}
W_m = \half \frac{e^2}{R}\left( \frac{v}{c} \right)^2 N^2 \left(
\frac{1}{2\pi} \sum_{i=1}^{N-1} V_{\varphi}(i\, \Delta\varphi)
\Delta\varphi \right) ,
\end{equation}
where $\Delta\varphi = 2\pi/N$. We note that the current in the
ring is $i = \frac{e\omega N}{2\pi}$ so we can write the above
expression in the form $W_m = \half L i^2$ where $L$, by
definition, is the self-inductance,
\begin{equation}\label{eq.self.induct.polyg}
L =\left( \frac{2\pi}{c} \right)^2 R\,  \frac{1}{2\pi}
\sum_{i=1}^{N-1} V_{\varphi}(i\, \Delta\varphi) \Delta\varphi ,
\end{equation}
of the polygon.

For all but the smallest values of $N$ it is useful to approximate
the sum with an integral. Using the trapezoidal rule
\cite{BKdahlquist&al},
\begin{equation}\label{eq.trapezoidal.rule}
\int_a^b f(x)\, \dfd x \approx \frac{b-a}{n} \left\{ \half f(a) +
\sum_{k=1}^{n-1} f\left( a+k\frac{b-a}{n} \right) + \half f(b)
\right\},
\end{equation}
we find that (taking $n=N-2$),
\begin{equation}\label{eq.approx.trapetz}
 \sum_{i=1}^{N-1} V_{\varphi}(i\, \Delta\varphi)
\Delta\varphi \approx  \int_{\frac{2\pi}{N}}^{2\pi-\frac{2\pi}{N}}
V_{\varphi}(\varphi)\, \dfd\varphi +\frac{\pi}{N} \left[
V_{\varphi}\left(\frac{2\pi}{N} \right) +
V_{\varphi}\left(2\pi-\frac{2\pi}{N}\right) \right].
\end{equation}
Using this and expanding the result in powers of $1/N$ one finds
the results,
\begin{equation}\label{eq.L.for.polyg.darwin}
L^D=\frac{4\pi}{c^2} R \left[ \ln\left(\frac{2N}{\pi} \right) - 1
\right]
\end{equation}
for the Darwin case (\ref{eq.V.phi.darw}), and,
\begin{equation}\label{eq.L.for.polyg.lorenz}
L^L=\frac{4\pi}{c^2} R \left[ \ln\left(\frac{2N}{\pi} \right) -
\frac{3}{2}\right]
\end{equation}
for the Lorenz case (\ref{eq.V.phi.lorenz}). The integral can be
done analytically. Terms in $1/N^4$ and higher were neglected in
the expansions. Maple \cite{BKheck} was used in these
calculations, as well as for most other calculations and plots of this review.

These results may be compared with the traditional result (Becker
\cite{BKbecker}),
\begin{equation}\label{eq.L.for.ring.trad.lor}
L_{\rm c}=\frac{4\pi}{c^2} R \left[ \ln\left(\frac{8R}{\lambda}
\right)  - \frac{7}{4} \right],
\end{equation}
for the self-inductance of a circular loop conductor used above in
(\ref{eq.self.induct.ring}).  Since $N=2\pi R/\delta$, where
$\delta$ is the distance between neighboring charges along the
circle, the logarithmic parts of the results agree if
$\delta=\lambda/2$. Equivalently the polygon must have $N=4\pi
R/\lambda$ electrons to agree with the logarithmic part of the
ring result, $\lambda$ being the cross sectional radius of the
ring.

For large $N$ the logarithmic part of the self-inductance should
be the dominating one. The contributions linear in $R$ are seen to
be different in all cases but do seem to have an order of
magnitude agreement.

There is a considerable literature on inductance calculations. The
books by Bueno and Assis \cite{BKbueno&assis} and by Grover
\cite{BKgrover} may be mentioned. Technical papers are
\cite{garrett,leferink,abakar&al}, and a few applications of ideas
concerning self-inductance can be found in
\cite{rammal&al,pendry&al,flores&ud}.


\end{document}